\documentclass[a4paper]{article}

\usepackage{graphicx}
\usepackage{natbib}
\usepackage{color}
\usepackage{amsmath}
\usepackage{latexsym}
\usepackage{amssymb}
\usepackage{authblk}

\newsavebox{\astrutbox}
\sbox{\astrutbox}{\rule[-5pt]{0pt}{20pt}}

\title{The streamwise turbulence intensity in the intermediate layer
  of turbulent pipe flow}

\author{J.C. VASSILICOS $^{1,2,4}$, J.-P. LAVAL$^{3,5}$, J.-M.
         FOUCAUT$^{2,5}$ and M. STANISLAS$^{2,5}$\\
\mbox{\vspace{1cm}}\\
$^1$ Department of Aeronautics, Imperial College London,\\
  London SW7 2AZ, United Kingdom\\
 $^2$ ECLille, LML, F-59650 Villeneuve d'Ascq, France \\
 $^3$ CNRS, UMR 8107, F-59650 Villeneuve d'Ascq, France  \\
 $^4$ USTL, LML, F-59650 Villeneuve d’Ascq, France \\
 $^5$ Univ Lille Nord de France, F-59000 Lille, France}

\begin{document}

\date{August 1, 2014}
\maketitle

\begin{abstract}

The spectral model of Perry, Henbest \& Chong (1986) predicts that the
integral length-scale varies very slowly with distance to the wall in
the intermediate layer. The only way for the integral length scale's
variation to be more realistic while keeping with the Townsend-Perry
attached eddy spectrum is to add a new wavenumber range to the model
at wavenumbers smaller than that spectrum. This necessary addition can
also account for the high Reynolds number outer peak of the turbulent
kinetic energy in the intermediate layer. An analytic expression is
obtained for this outer peak in agreement with extremely high Reynolds
number data by Hultmark, Vallikivi, Bailey \& Smits (2012, 2013). The
finding of Dallas, Vassilicos \& Hewitt (2009) that it is the eddy
turnover time and not the mean flow gradient which scales with
distance to the wall and skin friction velocity in the intermediate
layer implies, when combined with Townsend's (1976)
production-dissipation balance, that the mean flow gradient has an
outer peak at the same location as the turbulent kinetic energy. This
is seen in the data of Hultmark, Vallikivi, Bailey \& Smits
(2012, 2013). The same approach also predicts that the mean flow
gradient has a logarithmic decay at distances to the wall larger than
the position of the outer peak, a qualitative prediction which the
aforementioned data also support.

\end{abstract}

\section{Introduction}

Considering turbulent pipe/channel flows and turbulent boundary
layers, \cite{townsend76} developed his well-known attached-eddy model
to predict the profile with distance from the wall of the turbulent
kinetic energy. For wall distances much larger than the wall unit
$\delta_{\nu}$ and much smaller than, say, the pipe radius $\delta$,
which is the intermediate range where this model is operative, the
turbulent kinetic energy scales with the square of the wall friction
velocity $u_{\tau}$ and decreases logarithmically with distance to the
wall. However, measurements in turbulent boundary layers dating from
about twenty years ago (see \cite{fernholz&finley96}) as well as more
recent turbulent pipe flow measurements from the Princeton Superpipe
(\cite{morrisonetal04}, \cite{hultmarketal12}, \cite{hultmarketal13})
show that an outer peak appears in the mean square fluctuating
streamwise velocity at distances from the wall between about
$100\delta_{\nu}$ and $800\delta_{\nu}$ when the turbulent Reynolds
number $Re_{\tau} = \delta/\delta_{\nu}$ is larger than about
20 000. Such non-monotonic behaviour in regions where the mean flow is
monotonically increasing is hard to account for in current turbulence
models and theory, and inconceivable within the current framework of
Townsend's attached eddy model.

Starting with the spectral model of \cite{perryetal86} there have been
numerous developments and extensions of the attached eddy model (see
the review by \cite{smitsetal11} and references therein) but none has
accounted for the outer peak in turbulent kinetic energy. Here we
start from the observation (given in section 3) that the
\cite{perryetal86} attached edddy model has a basic shortcoming to do
with the integral length-scale it predicts. There is only one way to
repair this model without removing its attached eddy part, and this
way naturally leads to an outer peak in turbulent kinetic energy.

In section 2 we provide some basic background on the type of turbulent
pipe/channel flow considered in this paper and in section 3 we briefly
describe the Townsend-Perry attached eddy model and its consequences
on the integral scale. Section 4 is on the modification to the
Townsend-Perry attached eddy model that we are forced to implement to
remedy the integral scale problem. This section contains comparisons
between the predictions of this modified attached eddy model and the
Nano Scale Thermal Anemometry Probe (NSTAP) data obtained in the
Princeton Superpipe by \cite{hultmarketal12, hultmarketal13}. In
section 5 we explain how intermittency in wall shear stress
fluctuations could modify the attached-eddy $k_{1}^{-1}$ spectrum and
make is slightly steeper. In section 6 we predict that the mean flow
gradient must have an outer peak at the same distance from the wall
where the turbulent kinetic energy has its outer peak and report that
the data of \cite{hultmarketal12, hultmarketal13} show clear evidence
of this. We end the paper with a list of main conclusions in section
7. The words ``turbulence intensity'' appear in the title of this
paper because it is concerned primarily with the mean square
fluctuating streamwise velocity (sections 3 to 5) but also with the
streamwise mean flow (section 6).

%citation example \cite{laizet&lamballais09}

\section{Turbulent pipe/channel flow}

We consider a smooth pipe/channel that is long enough and a flow in it
operating at high enough Reynolds number and steadily driven by a
constant (in space and time) pressure gradient so that a turbulent
region exists far enough from the inlet where turbulence statistics
are independent of streamwise spatial coordinate $x$ and of time $t$.
The mean flow is $( \overline{u}, 0,0)$ and the fluctuating velocity
field is $(u', v', w')$ where $\overline{u}$ and $u'$ are along the
streamwise axis and $v'$ is parallel to the coordinate $y$ normal to
the wall.

The mean balance of forces along $x$, i.e. $-{1\over \rho}{d\over
  dx}\overline{P} = u_{\tau}^{2}/\delta$ where $\delta$ is the
half-width of the channel or the radius of the pipe, allows
determination of the skin friction velocity $u_{\tau}$ from
measurements of the mean pressure gradient $-{d\over dx}\overline{P}$
($\rho$ is the mass density of the fluid in the pipe/channel).

The wall unit is $\delta_{\nu} \equiv \nu/u_{\tau}$. It is well known
that if the Reynolds number is large enough then $\delta_{\nu} \ll
\delta$, e.g. see \cite{pope00}. In such flows, one often uses the
Reynolds number $Re_{\tau} \equiv \delta/\delta_{\nu}$ as reference
and high Reynolds number then trivially implies wide separation of
outer/inner length-scales and the existence of the intermediate layer
$\delta_{\nu} \ll y \ll \delta$ where $y$ is the wall-normal spatial
coordinate with $y=0$ at the wall.

For a given channel/pipe (i.e. a given $\delta$), a given fluid
(i.e. a given kinematic visosity $\nu$), a given driving pressure drop
(i.e. a given $u_{\tau}$) and at a given distance $y$ from the wall, a
streamwise wavenumber $k_1$ could be comparable to $1/\delta$, $1/y$,
$1/\eta$ (where $\eta \equiv (\nu^{3}/\epsilon)^{1/4}$ is the
Kolmogorov microscale which is a function of $y$ via its dependence on
kinetic energy dissipation rate per unit mass $\epsilon$) or
$1/\delta_{\nu}$.

The argument which shows that $\delta_{\nu}$ is smaller than $\eta$ is
based on the log-law of the wall and a direct balance between
production and dissipation which one classically expects to hold in
the $y$-region where the Prandtl-von K\'arm\'an law of the wall holds,
e.g. see \cite{townsend76}, \cite{pope00}. At extremely high
$Re_{\tau}$, this balance may be written as $u_{\tau}^{2} {d\over dy}
\overline{u} \approx \epsilon$ where we have replaced the Reynolds
stress by $u_{\tau}^{2}$, something which can be rigorously shown to
hold in the range $\delta_{\nu} \ll y \ll \delta$ as a consequence of
axial momentum balance in turbulent pipe/channel flows
%(to our knowledge not in turbulent boundary layers) 
under a very mild extra assumption, see section III in
\cite{dallasetal09}.
%The mean flow is $( \overline{u},
%0,0)$ and the fluctuating velocity field is $(u', v', w')$ where $v'$
%is in the direction normal to the wall.

This equilibrium argument implies that $\epsilon \sim u_{\tau}^{3}/y$
(assuming that the log-law ${d\over dy} \overline{u} \sim u_{\tau}/y$
holds) in $\delta_{\nu} \ll y \ll \delta$. It is now possible to
compare $\eta = (\nu^{3}/\epsilon)^{1/4}$ and $\delta_{\nu} =
\nu/u_{\tau}$ and it follows from $\delta_{\nu} \ll y$ that $1/\eta
\ll 1/\delta_{\nu}$ in the range $\delta_{\nu} \ll y \ll \delta$. It
is worth stressing that $1/\eta \ll 1/\delta_{\nu}$ and $\epsilon \sim
u_{\tau}^{3}/y$ were obtained on the basis that the range
$\delta_{\nu} \ll y \ll \delta$ is an equilibrium log-law range in a
pipe/channel flow. We revisit this assumption in section 6.

From the above arguments, where $y$ is much larger than $\delta_{\nu}$
but much smaller than $\delta$, the axis of wavenumbers $k_1$ is
marked by wavenumbers $1/\delta$, $1/y$, $1/\eta$ and $1/\delta_{\nu}$
in this increasing wavenumber order. This order of cross-over
wavenumbers is important in the spectral interpretation by
\cite{perryetal86} of Townsend's attached eddy hypothesis and its
consequences.

\section{The Townsend-Perry attached eddy model}

\cite{townsend76} assumed ``that the main, energy-containing motion is
made up of contributions from `attached' eddies with similar velocity
distributions'' and developed a physical space argument which led to 
\begin{equation}
{1\over 2} \overline{u'^{2}}(y)/u_{\tau}^{2} \approx
C_{s0} + C_{s1} \ln (\delta/y)
\label{eq:3.1}
\end{equation}
in the range $\delta_{\nu} \ll y \ll \delta$. The two constants
$C_{s0}$ and $C_{s1}$ are independent of $y$ and $Re_{\tau}$.

\cite{perryetal86} developed a spectral attached eddy model and argued
that where $\delta_{\nu} \ll y \ll \delta$, the streamwise energy
spectrum $E_{11}(k_{1}, y)$ has three distinct ranges:

\noindent
(i) $k_{1} < 1/\delta$ where $E_{11}(k_{1}) \approx u_{\tau}^{2}
\delta g_{o} (k_{1}\delta)$ which must be $E_{11}(k_{1}) \approx
C_{\infty} u_{\tau}^{2} \delta $ with a constant $C_{\infty}$ at small
enough wavenumbers;

\noindent
(ii) $1/\delta < k_{1} <1/y$ where $E_{11}(k_{1}) \approx C_{0}
u_{\tau}^{2} k_{1}^{-1}$ (the `attached eddy' range);

\noindent
(iii) $1/y <k_{1}$ where $E_{11}(k_{1})$ has the Kolmogorov form
$E_{11} (k_{1}, y) \sim \epsilon^{2/3} k_{1}^{-5/3} g_{K} (k_{1}y,
k_{1}\eta)$, see \cite{pope00}, \cite{frisch95}.\\

By integration of $E_{11}(k_{1})$ they obtained for $\delta_{\nu} \ll
y \ll \delta$
\begin{equation}
{1\over 2} \overline{u'^{2}}(y)/u_{\tau}^{2} \approx
C_{\infty} + C_{0} \ln (\delta/y)
\label{eq:3.2}
\end{equation}
where the constants $C_{\infty}$ and $C_{0}$ are independent of $y$
and $Re_{\tau}$. Application of a strict matching condition for the
energy spectra at $k_{1} = 1/\delta$ gives $C_{0}=C_{\infty}$ but this
is of course not necessary. In fact, the constant $C_{\infty}$ in
equation (\ref{eq:3.2}) is not the same as the constant $C_{\infty}$ in the
spectral model if we allow for the wavenumber dependency of the outer
function $g_{o} (k_{1}y)$ and for the fact that this constant has a
small contribution from the high wavenumber Kolmogorov range
(iii). The detail of this Kolmogorov contribution has been neglected
in equation (\ref{eq:3.2}) as it only adds a term proportional to
$1-y_{+}^{-1/2}$ to the right hand side ($y_{+} \equiv
y/\delta_{\nu}$) which is of little effect in the considered range.

A consequence of the \cite{perryetal86} model is that the integral
scale $L_{11}$ is proportional to $\delta$ and very weakly dependent
on $y$ in the intermediate layer $\delta_{\nu} \ll y \ll \delta$. This
follows from $\pi E_{11}(k_{1}=0 , y) = \overline{u'^{2}}(y)
L_{11}(y)$ (e.g. see \cite{tennekes&lumley72}) which leads to
\begin{equation}
L_{11} (y) \approx {\pi C_{\infty} \delta \over C_{\infty} + C_{0} \ln
  (\delta/y)}
\label{eq:3.3}
\end{equation}
where $\delta_{\nu} \ll y \ll \delta$. However, one expects that
$L_{11}$ may depend on $y$ much more steeply. For example, the
turbulent boundary layer measurements of \cite{tomkins&adrian03}
suggest that $L_{11} \sim y$.

The only way for the Towsend-Perry attached eddy wavenuber range to be
viable, i.e. the only way to have an integral scale which depends more
substantially on $y$ while keeping with the Townsend-Perry attached
eddy wavenumber range (where, in particular, the constant $C_0$ is
independent of $y$ and $Re_{\tau}$) is to modify the model of
\cite{perryetal86} by inserting a fourth range to $E_{11}(k_{1})$
between the very low-wavenumber range where $E_{11}(k_{1}) \approx
C_{\infty} u_{\tau}^{2} \delta$ and the `attached eddy' range. We
develop such a model in the following section.

\section{A modified Townsend-Perry attached eddy model}

We now consider a model of the energy spectrum $E_{11}(k_{1},y)$ with
the following four ranges

\noindent
(i) $k_{1} < 1/\delta_{\infty}$ where $E_{11}(k_{1}) \approx
C_{\infty} u_{\tau}^{2} \delta $ with a constant $C_{\infty}$
independent of wavenumber;

\noindent
(ii) $1/\delta_{\infty} < k_{1} <1/\delta_{*}$ where $E_{11}(k_{1})
\approx C_{1} u_{\tau}^{2} \delta (k_{1}\delta)^{-m}$ where $0<m<1$
and $C_{1}$ is also a constant independent of wavenumber;

\noindent
(iii) $1/\delta_{*} < k_{1} <1/y$ where $E_{11}(k_{1}) \approx C_{0}
u_{\tau}^{2} k_{1}^{-1}$ where $C_0$ is a constant independent of
wavenumber, $y$ and $Re_{\tau}$ (the `attached eddy' range);

\noindent
(iv) $1/y <k_{1}$ where $E_{11}(k_{1})$ has the Kolmogorov form
$E_{11} (k_{1}, y) \sim \epsilon^{2/3} k_{1}^{-5/3} g_{K} (k_{1}y,
k_{1}\eta)$.

Note the presence of the two new length-scales $\delta_{\infty}$ and
$\delta_{*}$. The only physics that we impose is the expectation that
this range grows as the position $y$ where $E_{11}(k_{1}, y)$ is
evaluated approaches the wall and distances itself from the centre of
the pipe within $\delta_{\nu} \ll y\ll \delta$. The range
$(1/\delta_{*})/(1/\delta_{\infty}) = \delta_{\infty}/\delta_{*}$ can
only depend on $y$, $\delta$, $\nu$ and $u_{\tau}$. Without loss of
generality, it is therefore a function of $y/\delta$ and $Re_{\tau}$
or, equivalently, $y_{+}$ and $Re_{\tau}$.  At fixed $Re_{\tau}$,
$\delta_{\infty}/\delta_{*}$ must be a decreasing function of
$y/\delta$ and also a decreasing function of $y_{+}$. At fixed
$y/\delta$, $\delta_{\infty}/\delta_{*}$ must be a decreasing function
of $Re_{\tau}$ as this implies that $y_{+}$ increases. And at fixed
$y_{+}$, $\delta_{\infty}/\delta_{*}$ must be an increasing function
of $Re_{\tau}$ as this means that $y/\delta$ decreases.

An arbitrary but not impossible functional
dependence is
\begin{equation} 
\delta_{\infty}/\delta_{*}\approx A \; (y/\delta)^{-p} Re_{\tau}^{-q}
\approx A y_{+}^{-p} Re_{\tau}^{p-q}
\label{eq:4.1}
\end{equation}
where $A$ is a dimensionless constant. The qualitative physics which
we described in the previous paragraph impose $p,q>0$ and $p>q$. We
adopt equation (\ref{eq:4.1}) indicatively in what follows as the aim of this
work is to show the possibilities which open up with the adoption of
the extra wavenumber range $1/\delta_{\infty} < k_{1} < \delta_{*}$
for the purpose of reconciling the Townsend-Perry attached eddy
hypothesis with a more realistic integral length-scale. We limit the
values of the exponents $p$ and $q$ to $p,q>0$ and $p>q$ without
further constraints.

Matching of the energy spectral forms at $k_{1} \approx
1/\delta_{\infty}$ gives $C_{\infty} = C_{1}
(\delta/\delta_{\infty})^{-m}$ and at $k_{1} \approx 1/\delta_{*}$
gives $C_{1} = C_{0} (\delta/\delta_{*})^{m-1}$. It is not strictly
necessary to impose these matching conditions as they unnecessarily
restrict the cross-over forms of the energy spectra, but they do
indicate that we need an expression for $\delta_{*}/\delta$ if we are
to proceed with or without them. Given that in all generality,
$\delta_{*}/\delta$ is a function of $y/\delta$ and $Re_{\tau}$, we
again assume a power-law form
\begin{equation}
\delta_{*}/\delta = B \; (y/\delta)^{\alpha} Re_{\tau}^{\beta} 
\label{eq:4.2}
\end{equation}
where, like $A$, $B$ is a dimensionless constant.

There are also two requirements for the viability of our spectra:
$y\ll \delta_{*}$ and $\delta_{*} < \delta_{\infty}$. The former is
met provided that $\beta \ge \alpha -1$ for $y\gg \delta_{\nu}$. The
latter is met if $y < y_{*} \equiv \delta A^{1/p} Re_{\tau}^{-q/p}$.

We therefore adopt the new range (ii) for $y < y_{*}$ but keep the
\cite{perryetal86} model unaltered for $y > y_{*}$.  Their model can
indeed remain unaltered if $\delta_{\infty}=\delta_{*} = \delta$ at $y
\ge y_{*} = \delta A^{1/p} Re_{\tau}^{-q/p}$. The continuous passage
from (\ref{eq:4.1}) and (\ref{eq:4.2}) to $\delta_{\infty}=\delta_{*} = \delta$ requires
$\beta = \alpha q/p$ and $BA^{\alpha/p}=1$.

By integration of $E_{11}(k_{1})$ we obtain for $\delta_{\nu} \ll
y \le y_{*}$
\begin{equation}
{1\over 2} \overline{u'^{2}}(y)/u_{\tau}^{2} \approx
C_{s0} - C_{s1}\ln (\delta/y)
-C_{s2} (y/\delta)^{p(1-m)}Re_{\tau}^{q(1-m)}
\label{eq:4.3}
\end{equation}
where $C_{s0} = {C_{0}\over 1-m} + C_{0} \ln B + C_{0}\alpha {q\over
  p} \ln Re_{\tau} \;\;$, $C_{s1} = C_{0} (\alpha-1)$ and $C_{s2} =
{mC_{0} A^{m-1}\over 1-m}$. (Note that $C_{s0}$ is a weak function of
$Re_{\tau}$ whereas $C_{s1}$ and $C_{s2}$ are independent of
$Re_{\tau}$.) These new constants have been calculated by taking into
account the perhaps over-constraining matching conditions $C_{\infty}
= C_{1} (\delta/\delta_{\infty})^{-m}$ and $C_{1} = C_{0}
(\delta/\delta_{*})^{m-1}$.

The integral length scale is now
\begin{equation}
L_{11}/\delta= \pi \, C_{0} \, A^{m} \, B \; (y/\delta)^{\alpha -pm}
Re_{\tau}^{\beta-qm}/(\overline{u'^{2}}(y)/u_{\tau}^{2})
\label{eq:4.4}
\end{equation}
clearly more strongly dependent on $y$ than in equation (\ref{eq:3.3}).

Equation (\ref{eq:4.3}) can be compared with the Townsend-Perry form which
remains valid here for $y_{*}\le y \ll \delta$ and which is (taking
$C_{\infty} = C_{0}$)
\begin{equation}
{1\over 2} \overline{u'^{2}}(y)/u_{\tau}^{2} \approx C_{0}+ C_{0}\ln
(\delta/y)
\label{eq:4.5}
\end{equation}

The two profiles (\ref{eq:4.3}) and (\ref{eq:4.5}) match at $y=y_{*} \equiv \delta
A^{1/p} Re_{\tau}^{-q/p}$ and so do also the integral length-scale
forms (\ref{eq:4.4}) and (\ref{eq:3.3}) if $C_{\infty} = C_{0}$. Our approach does not
modify the Townsend-Perry form of $L_{11}$ at large distances from the
wall, i.e at $y>y_{*}$, but it does return a siginificant dependence
of $L_{11}$ on $y$ which, however, is arbitrarily set by equations
(\ref{eq:4.1}) and (\ref{eq:4.2}). Even so, the possibility is now open for a stronger
dependence of $L_{11}$ on $y$. This possibility has been opened by the
adoption of an extra wavenumber range $1/\delta_{\infty} < k_{1} <
1/\delta_{*}$ which, in turn, returns a form of the
$\overline{u'^{2}}(y)$ profile which allows for a maximum value (a
peak) inside the intermediate region $\delta_{\nu} \ll y\ll
\delta$. No such peak is allowed by the Townsend-Perry forms (\ref{eq:3.1}) and
(\ref{eq:3.2}) although such a peak has been observed in measurements of both
turbulent boundary layers and turbulent pipe flows over the past 20
years or so, see \cite{fernholz&finley96}, \cite{morrisonetal04},
\cite{hultmarketal12}, \cite{hultmarketal13}.

Straightforward analysis of (\ref{eq:4.3}) shows that a maximum streamwise
turbulence intensity does exist in the range $\delta_{\nu}\ll y \ll
\delta$ if $0< \alpha -1 < pm$ (i.e. if $C_{s1} >0$ and $\alpha<
pm+1$) and that the position $y_{peak}$ of this maximum is
\begin{equation}
y_{peak}/\delta \sim Re_{\tau}^{-q/p}
\label{eq:4.6}
\end{equation}
which decreases with increasing $Re_{\tau}$ and , equivalently, 
\begin{equation}
y_{peak}/\delta_{\nu} \sim Re_{\tau}^{1-q/p}
\label{eq:4.7}
\end{equation}
which increases with increasing $Re_{\tau}$ as $q<p$. It also follows
from (\ref{eq:4.3}) that 
\begin{equation}
{d\over d\ln Re_{\tau}}\left( {1\over 2} 
  \overline{u'^{2}}(y_{peak})/u_{\tau}^{2} \right) \approx C_{0} (\alpha p/q
  -\alpha q/p + q/p) >0.
\label{eq:4.8}
\end{equation}
The maximum value of $\overline{u'^{2}}(y)/u_{\tau}^{2}$ at
$y=y_{peak}$ therefore grows logarithmically with increasing
$Re_{\tau}$.

We now compare our functional dependence of ${1\over 2}
\overline{u'^{2}}(y)/u_{\tau}^{2}$ on $y$ and $Re_{\tau}$ with smooth
wall turbulent pipe flow data obtained recently with a new Nano Scale
Thermal Anemometry Probe (NSTAP) as reported by \cite{hultmarketal12,
  hultmarketal13}. Below we refer to this data as NSTAP Superpipe
data.

We start by fitting the data with 
(\ref{eq:4.5}) in the range $y_{*} < y \ll \delta$ and 
\begin{equation}
{1\over 2} \overline{u'^{2}}(y)/u_{\tau}^{2} \approx
C_{s0} - C_{s1}\ln (\delta/y)
-C_{s2} (y/\delta)^{d_{1}}Re_{\tau}^{d_{2}}
\label{eq:4.9}
\end{equation}
instead of (\ref{eq:4.3}) in the range $\delta_{\nu} \ll y < y_{*}$ where
$y_{*} = \delta Re_{\tau}^{-d_{2}/d_{1}}$. This is a model where we
ignore the various matching conditions which led to (\ref{eq:4.3}) with the
specific relations between $C_{s0}$, $C_{s1}$ and $C_{s2}$ and
the parameter $C_{0}$, $m$, $p$, $q$, $A$, $\alpha$ and $Re_{\tau}$.
It is also a model where we just set $A=1$, $d_{1} = p(1-m)$ and
$d_{2} = q (1-m)$ so that $y_{*} = \delta
Re_{\tau}^{-d_{2}/d_{1}}$. In figure 1 we show the result of this fit
against the NSTAP Superpipe data and in figure 2 we show the fitting
values of $C_{s0}$, $C_{s1}$, $C_{s2}$ and $d_1$ and $d_2$ and
their dependence on $Re_{\tau}$ in a lin-log plot.

%\begin{figure}[!h]
\begin{figure}
\centering 
\vspace{1cm}
\begin{minipage}{0.49\columnwidth}
\includegraphics[width=\columnwidth]{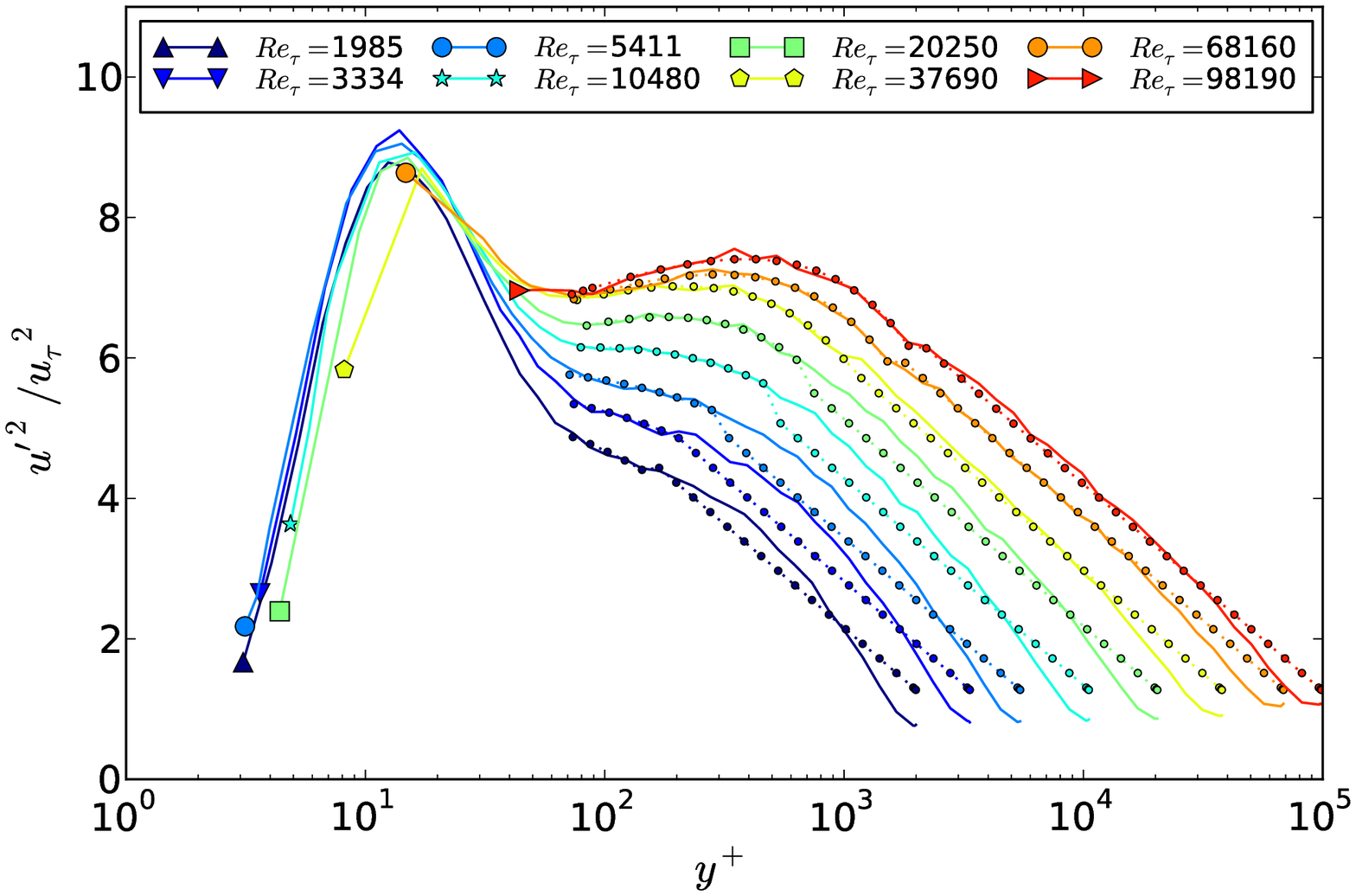}
\end{minipage}
\hfill
\begin{minipage}{0.49\columnwidth}
\includegraphics[width=\columnwidth]{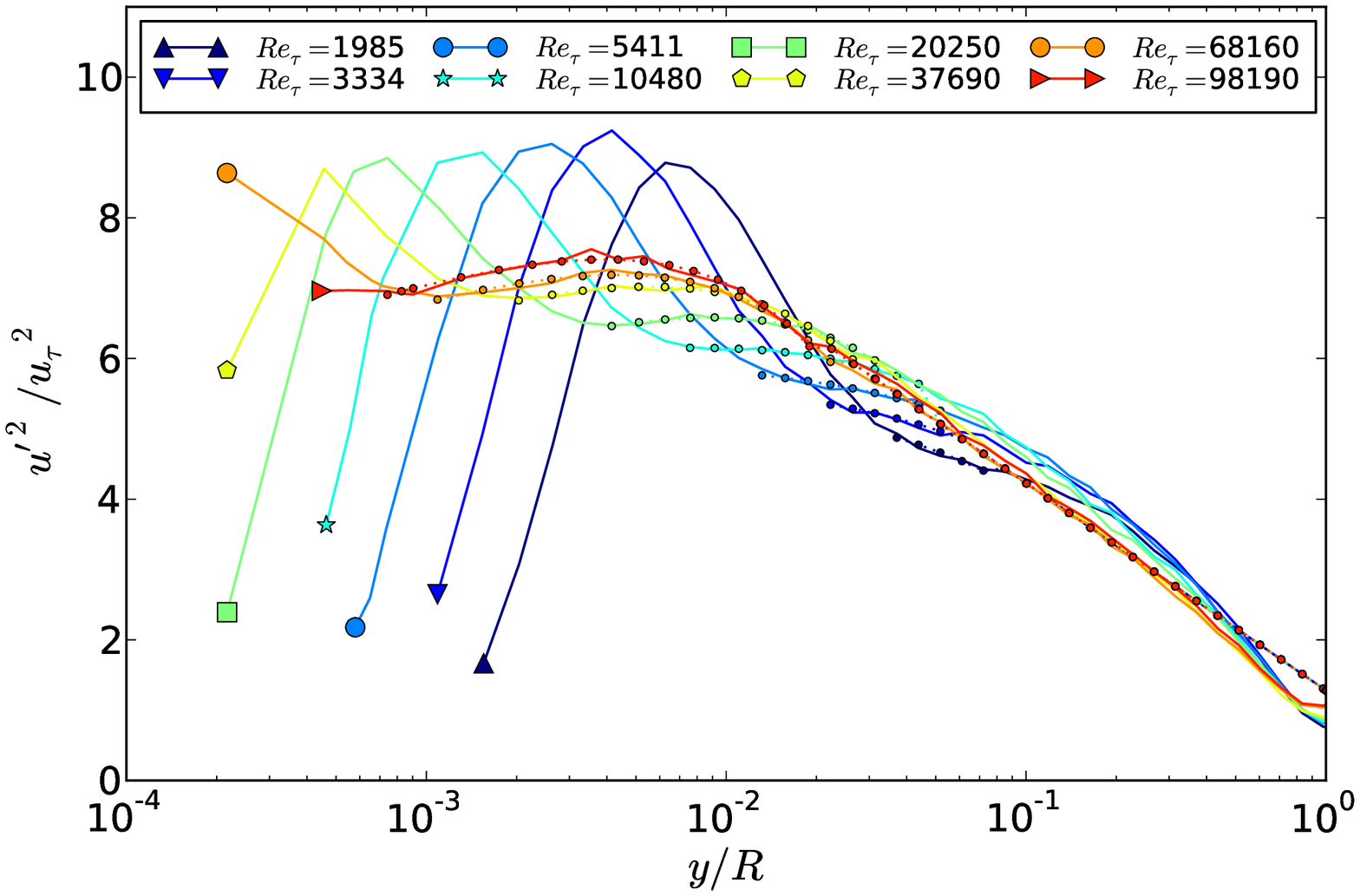}
\end{minipage}
\caption{Plots of $\overline{u'^{2}}(y)/u_{\tau}^{2}$ versus $y_+$
  (left) and $y/\delta$ (right) obtained from the NSTAP Superpipe data
  of \cite{hultmarketal12, hultmarketal13} for different values of
  $Re_{\tau}$. The circles are calculated from equations (\ref{eq:4.5}) and
  (\ref{eq:4.9}) with $C_{0} = 1.28$, $y_{*} = \delta Re_{\tau}^{-d_{2}/d_{1}}$
  for all Reynolds numbers and the values of $d_1$ and $d_2$ and the
  constants in (\ref{eq:4.9}) given in figure \ref{fig:2}.}
\label{fig:1}
\end{figure}

First note in figure \ref{fig:1} the clear presence when $Re_{\tau}$
is larger than about $20\,000$ of a logarithmic region at the higher
$y$-values in agreement with the Townsend-Perry equation
(\ref{eq:4.5}) which fits it quite well (the fit is much better if we
allow $C_{\infty}$ to be different from $C_0$ as in equation
(\ref{eq:3.2})). This was of course already noted by
\cite{hultmarketal12, hultmarketal13}. Secondly note the gradual
development as $Re_{\tau}$ increases of a peak of turbulence intensity
inside the intermediate region $\delta_{\nu}\ll y\ll \delta$. This
outer peak is distinct from the well known near-wall peak at $y_{+}
\approx 15$ and starts appearing clearly at $Re_{\tau}$ larger than
about $20\,000$. Of course this was also noted in
\cite{hultmarketal12, hultmarketal13} who pointed out that the
position $y_{peak}$ of the outer peak depends on Reynolds number as
$y_{peak}/\delta_{\nu} \approx 0.23 Re_{\tau}^{0.67}$. In terms of our
model this means $d_{2}/d_{1} = q/p \approx 1/3$. As predicted by the
minimal physics instilled in our model (see the paragraph containing
equation (\ref{eq:4.1}) and the paragraph preceding it)
$y_{peak}/\delta$ decreases and $y_{peak}/\delta_{\nu}$ increases with
increasing $Re_{\tau}$ (see figure \ref{fig:1}). As also predicted by
the minimal physics of our model, the value of
$\overline{u'^{2}}/u_{\tau}^{2}$ at the outer peak slowly increases
with increasing $Re_{\tau}$ and the fits in figure 1 which we discuss
in the following paragraph indicate that this increase is indeed only
logarithmic as in equation (\ref{eq:4.8}).

\begin{figure}
\centering 
\vspace{1cm}
\includegraphics[width=0.8\columnwidth]{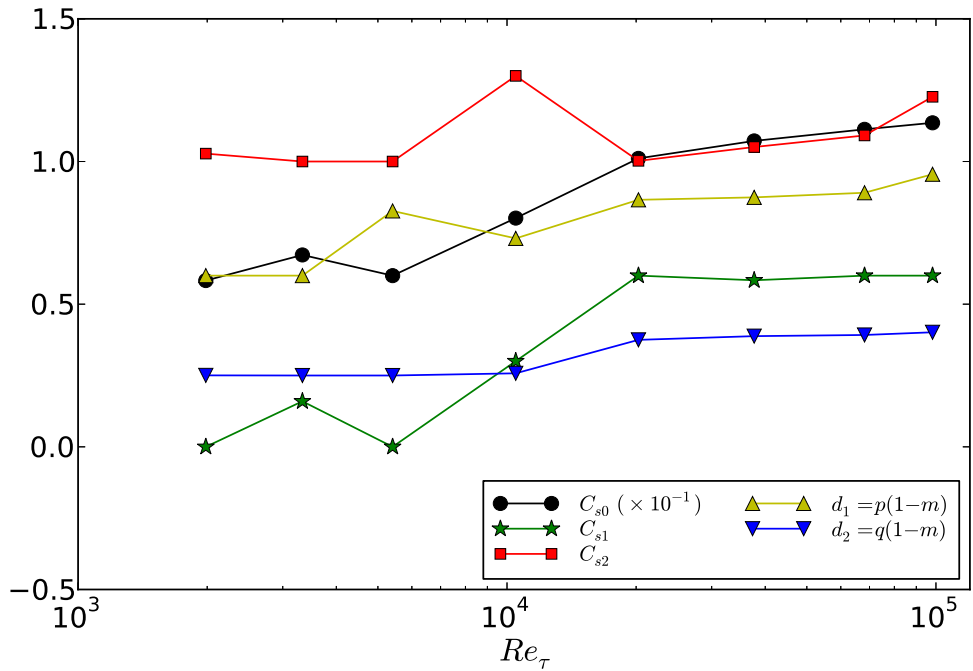}
\caption{Model parameters $C_{s0}$, $C_{s1}$, $C_{s2}$, $d_1$ and
$d_2$ appearing in equation (\ref{eq:4.9}). Plotted as functions
of $Re_{\tau}$.}
\label{fig:2}
\end{figure}

The point $y=y_{*}$ is clearly seen in figure \ref{fig:1} because we
did not adopt matching conditions to ensure a continuous passage from
(\ref{eq:4.9}) to (\ref{eq:4.5}). Nevertheless the new equation
(\ref{eq:4.9}) returns a satisfactory fit of the outer peak, including
its shape, intensity and location.  In figure \ref{fig:2} we plot the
Reynolds number dependence of the constants $C_{s0}$, $C_{s1}$ and
$C_{s2}$, $d_1$ and $d_2$ involved in these fits.  Note how all the
parameters $C_{s0}$, $C_{s1}$, $C_{s2}$, $d_1$ and $d_2$ do not
deviate much from a constant value for $Re_{\tau}$ larger than about
$20\,000$.

\begin{figure}
\centering 
\vspace{1cm}
\begin{minipage}{0.49\columnwidth}
\includegraphics[width=\columnwidth]{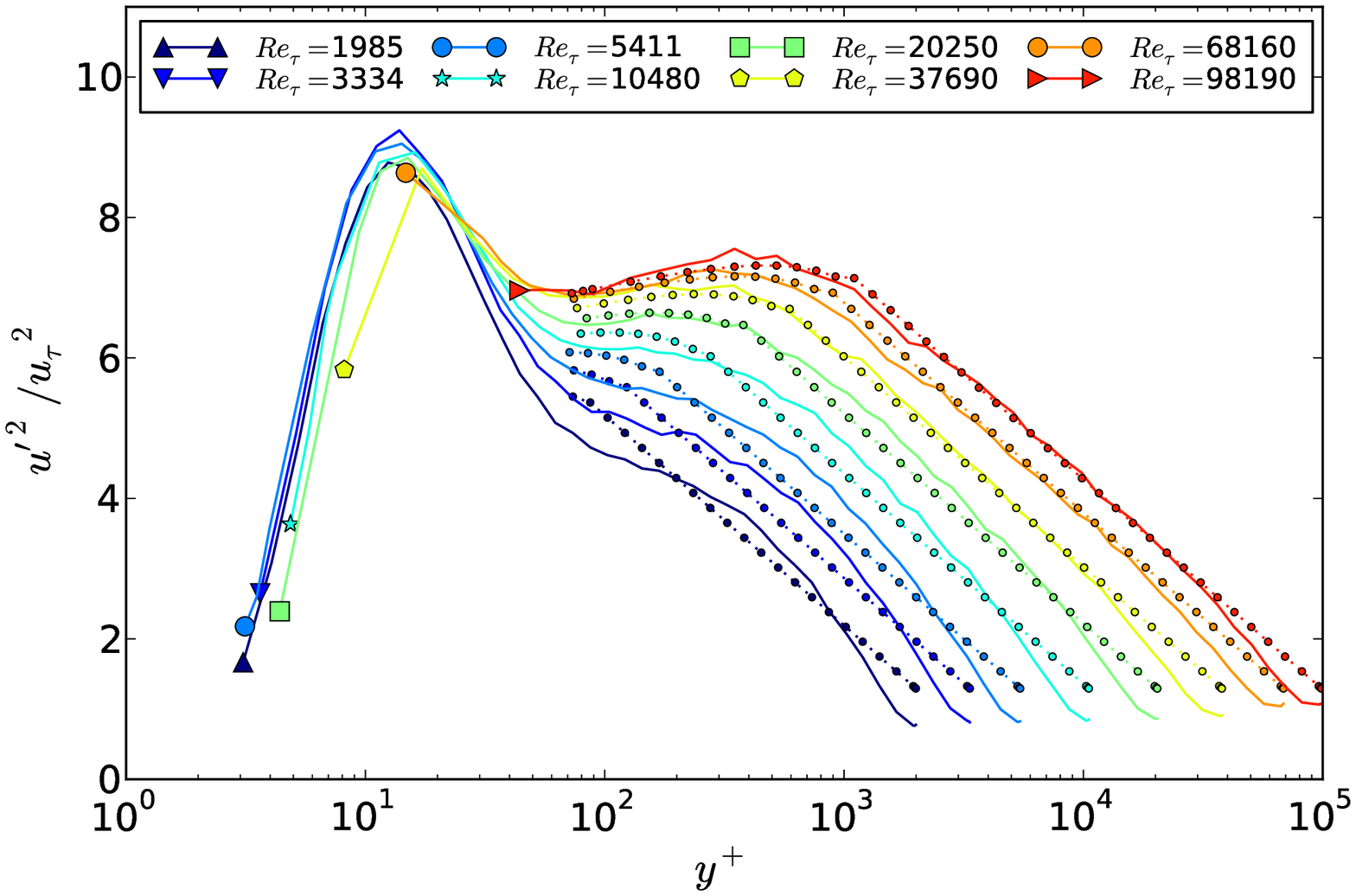}
\end{minipage}
\hfill
\begin{minipage}{0.49\columnwidth}
\includegraphics[width=\columnwidth]{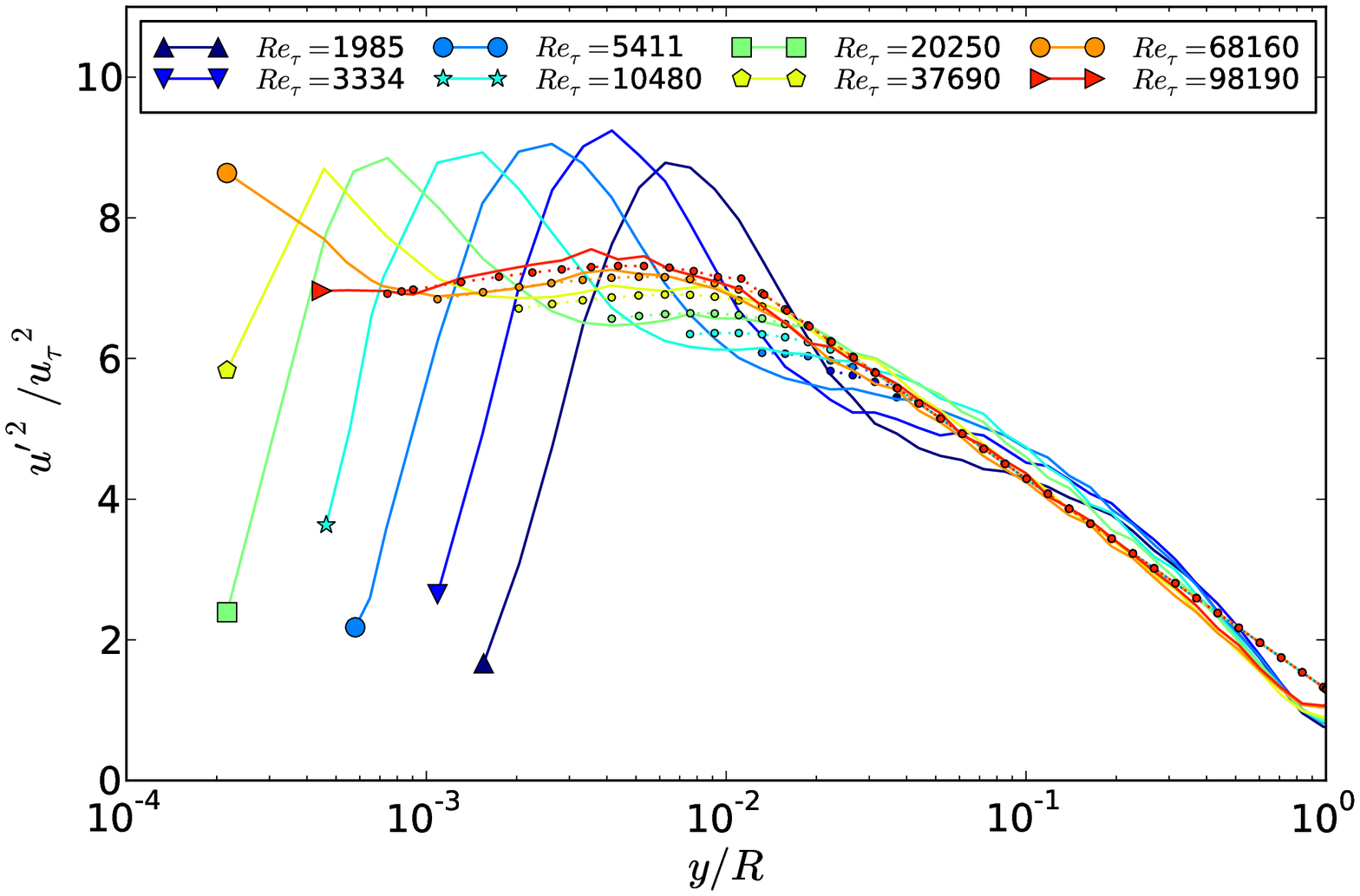}
\end{minipage}
\caption{Plots of $\overline{u'^{2}}(y)/u_{\tau}^{2}$ versus $y_+$
  (left) and $y/\delta$ (right) obtained from the NSTAP Superpipe data
  of \cite{hultmarketal12, hultmarketal13} for different values of
  $Re_{\tau}$. The circles are calculated for all Reynolds numbers
  from equations (\ref{eq:4.5}) and (\ref{eq:4.3}) with $y_{*} = \delta A^{1/p}
  Re_{\tau}^{-q/p}$ and $A=0.2$, $C_{0} = 1.28$, $m=0.37$, $q=0.79$,
  $p=2.38$ and $\alpha = 1.21$.}
\label{fig:3}
\end{figure}

In figure \ref{fig:3} we fit the NSTAP Superpipe data with (\ref{eq:4.5}) in the
range $y_{*} < y \ll \delta$ and (\ref{eq:4.3}) in the range $\delta_{\nu} \ll y
< y_{*}$ where $y_{*} = \delta A^{1/p}Re_{\tau}^{-q/p}$ and with $C_{s0}$, 
$C_{s1}$ and $C_{s2}$ given by 
\begin{eqnarray}
C_{s0} &=& {C_{0}\over 1-m} + C_{0} \ln B + C_{0}\alpha {q\over p}
\ln Re_{\tau}, \label{eq:4.10} \\
C_{s1} &=& C_{0} (\alpha-1), \label{eq:4.11} \\
C_{s2} &=& {mC_{0} A^{m-1}\over 1-m} \label{eq:4.12}
\end{eqnarray}

where $B= A^{\alpha/p}$ as obtained above in the text between
equations (\ref{eq:4.2}) and (\ref{eq:4.3}). The fits in figure 3 are obtained for
$A=0.2$, $C_{0} = 1.28$, $m=0.37$, $q=0.79$, $p=2.38$ and $\alpha =
1.21$. It works rather well, though not perfectly, for $Re_{\tau}$
larger than about $30 000$. Note that we did not optimise the choice
of our fitting parameters to obtain the best possible fit. As things
stand, equation (\ref{eq:4.9}) fits better the outer peak than equation (\ref{eq:4.3})
with (\ref{eq:4.10}), (\ref{eq:4.11}), (\ref{eq:4.12}) and $B= A^{\alpha/p}$. However, as of
course expected, the latter over-matched model returns a continuous
transition to (\ref{eq:4.5}) at $y=y_*$. Note that $y_{peak} \approx 0.45 \,
y_{*}$ (from $y_{peak}/\delta_{\nu} \approx 0.23 \, Re_{\tau}^{0.67}$ and
$y_{*} = \delta A^{1/p}Re_{\tau}^{-q/p}$).

Indicatively and only for illustrative purposes, we mention that the
fits in figure \ref{fig:3} correspond, approximately (we have rounded
off the exponents to make them look like fractions without any
intention to suggest a deeper level of theory), to
$\delta_{\infty}/\delta_{*} \approx 0.2 (y/\delta)^{-7/3}
Re_{\tau}^{-4/5}$ and $\delta_{*} \approx 0.44 \delta (y/\delta)^{6/5}
Re_{\tau}^{2/5}$ given that $\beta = \alpha q/p$. The model leading to
these particular fits also effectively assumes that the longitudinal
spectra in the region $\delta_{\nu} \ll y < y_{*} \approx 0.5 \delta
Re_{\tau}^{-1/3}$ have a range of wavenumbers $1/\delta_{\infty} <
k_{1} < 1/\delta_{*}$ which are lower than the usual attached eddy
ones and where $E_{11}(k_{1}) \approx {2\over 3} u_{\tau}^{2} y
Re_{\tau}^{1/3} (k_{1}\delta)^{-1/3} = {2\over 3} u_{\tau}^{2} y
(k_{1}\delta_{\nu})^{-1/3}$. Note the presence of both $y$ and
$\delta_{\nu}$ in these particularly low-wavenumber spectra. Note also
that $\delta_{*} < 0.2 \delta$ and $\delta_{\infty} > 5\delta/100$
given that $y < y_{*} \approx 0.5 \delta Re_{\tau}^{-1/3}$. Finally,
$y_{*} >15\delta_{\nu}$ as long as $Re_{\tau}> 165$.

In the region $y_{*} \approx 0.5 \delta Re_{\tau}^{-1/3} < y \ll
\delta$ no such spectral range exists; only the attached eddy form
$E_{11} \approx 1.28 u_{\tau}^{2} k_{1}^{-1}$ is present in the usual
range $1/\delta < k_{1} < 1/y$. The constant $C_{0}=1.28$ is the one
used to fit the data in both figures \ref{fig:3} and \ref{fig:1}.

\begin{figure}
\centering 
\vspace{1cm}
\includegraphics[width=0.8\columnwidth]{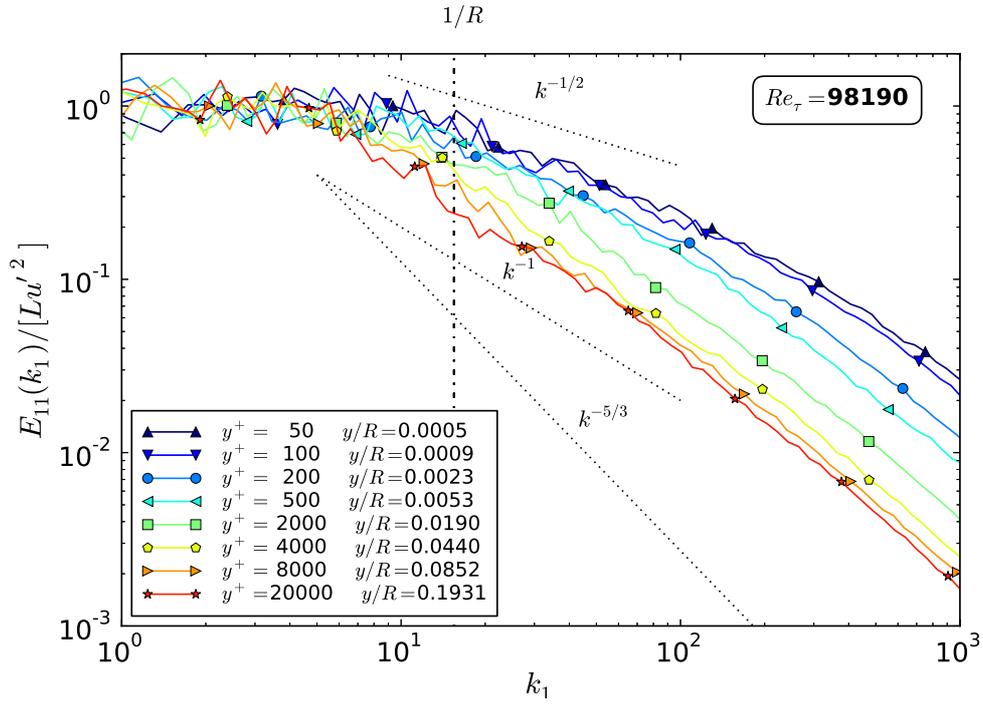}
\caption{NSTAP Superpipe energy spectra $E_{11} (k_{1},y)$ at various
  distances from the wall for $Re_{\tau} = 98\,190$.  At this Reynolds
  number, $y_{*}/\delta_{\nu} \approx 2130$. The spectra are
  normalised by $\overline{u'^{2}}(y) L_{11} (y)$ where $L_{11} (y)$
  are the integral scales obtained from these spectra.}
\label{fig:4}
\end{figure}

Figure \ref{fig:4} shows spectra plotted indicatively as wavenumber spectra at
many distances from the wall for a value of $Re_{\tau}$ equal to $98
190$ and $y_{*}/\delta_{\nu} \approx 2130$. These spectra are really
frequency spectra as we cannot expect the Taylor hypothesis to be
accurate enough at the lower wavenumbers and at the closer positions
to the wall. With this serious caveat firmly in mind it is
nevertheless intriguing to see in figure \ref{fig:4} that very high Reynolds
number spectra do indeed have an extra low-frequency range at $y <
y_{*}$ where the spectrum is much shallower than $k_{1}^{-1}$ yet not
constant; and that this range is absent at higher positions from the
wall where $y>y_{*}$.
At distances $y$ from the wall larger than $y_{*}$ one sees a
spectral wavenumber dependence which is close to $k_{1}^{-1}$ (perhaps
a little steeper) between a very low-wavenumber constant spectrum and
a very high-wavenumber spectrum which is much steeper than
$k_{1}^{-1}$, perhaps close to $k_{1}^{-5/3}$. Even the deviation from
the $k_{1}^{-1}$ spectrum which makes it look a little steeper could
be a frequency domain signature which does not quite correspond to
$k_{1}^{-1}$ because of Taylor hypothesis failure, see
\cite{delalamo&jimenez09} but also \cite{rosenberg13}.

\begin{figure}
\centering 
\vspace{1cm}
\includegraphics[width=0.8\columnwidth]{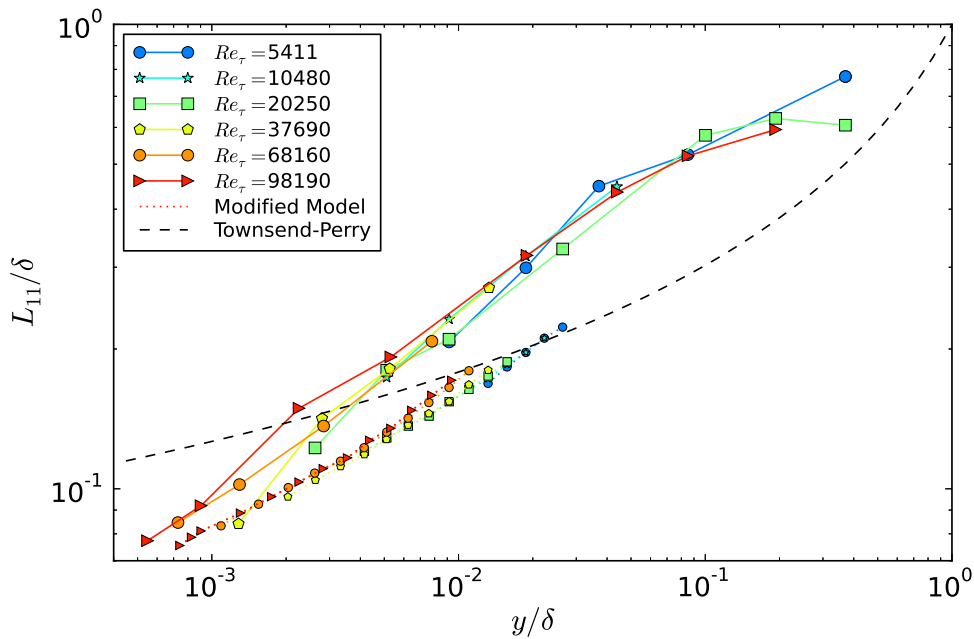}
\caption{Normalised integral scales $L_{11}/\delta$ obtained from
  NSTAP Superpipe energy spectra plotted versus $y/\delta$ for various
  Reynolds numbers. Also plotted are the Townsend-Perry and our
  modified model's prediction for $L_{11}/\delta$.}
\label{fig:5}
\end{figure}

Our initial motivation for modifying the \cite{perryetal86} model and
adding an extra spectral range to it was the $y$-dependence of the
integral scale. The values of the exponents $\alpha$, $q$, $p$ and $m$
used in the fits of figure \ref{fig:3} combined with the constraint $\beta
=\alpha q/p$ are such that $L_{11}/\delta \sim (y/\delta)^{1/3}
Re_{\tau}^{0.1}$ if we neglect the logarithmic dependence of
$\overline{u'^{2}}(y)/u_{\tau}^{2}$ in (\ref{eq:4.4}). In figure \ref{fig:5} we plot
$L_{11}/\delta$ versus $y/\delta$ as obtained from the lowest
frequencies of the NSTAP Superpipe spectra (see for example figure 4)
for different Reynolds numbers. Again, the integral scales plotted in
figure 5 should be taken with much caution and only very indicatively
as they are really integral time scales and the Taylor hypothesis
cannot be invoked at these low frequencies.  In that same figure we
nevertheless plot the Townsend-Perry formula (\ref{eq:3.3}) where $C_{\infty} =
C_{0}$ as per the fitting constants for figure \ref{fig:3}  (i.e. $L_{11} \approx
{\pi \delta\over 1+ \ln (\delta/y)}$) and formula (\ref{eq:4.4}). In (\ref{eq:4.4}) we
used the fitting constants that we also used for the fits in figure
\ref{fig:3}. Note that (\ref{eq:4.4}) is defined for $y$ in the range $\delta_{\nu}\ll y
<y_{*} = 0.5 \delta Re_{\tau}^{-1/3}$ and that, even in the modified
model, $L_{11}$ is given by (\ref{eq:3.3}) in the range $y_{*}\ll y < \delta$.
The points in figure \ref{fig:5} where the modified model curves meet the
Townsend-Perry curve are at $y=y_*$ for the different $Re_{\tau}$. It
is clear that the modified model succeeds in steepening the
$y$-dependence of $L_{11}$ in the range $\delta_{\nu}\ll y <y_{*}$ and
that it keeps the original $y$-dependence of $L_{11}$ in the range
$y_{*}\ll y < \delta$.
It is also clear, though, that formulae (\ref{eq:4.4}) and (\ref{eq:3.3}) do not match
the NSTAP Superpipe integral scales well with the fitting constants
used for figure \ref{fig:3}. We repeat that the integral scales obtained from
the NSTAP Superpipe data are really integral time scales and it is not
clear that they should be proportional to $L_{11}$. If such a
proportionality could be established, however, then the data would
indicate that $L_{11}/\delta \sim (y/\delta)^{1/3}$ for all Reynolds
numbers in some agreement with our modified model's $L_{11}/\delta
\sim (y/\delta)^{1/3} Re_{\tau}^{0.1}$, but the constants of
proportionality are different.

Finally, we draw attention to the fact that the integral scale
$L_{11}$ is not proportional to $y$ in the range $\delta_{\nu} \ll y
\ll \delta$ as one might have expected (see \cite{tomkins&adrian03}
who found several spanwise length scales, including $L_{11}$, to be
proportional to $y$ in a turbulent boundary layer).

\section{Intermittent attached eddies}

We now address the possibility brought up by experimental results such
as figure \ref{fig:4} that, in the appropriate Townsend-Perry attached eddy
range of wavenumbers, the energy spectra may not scale as $k_{1}^{-1}$
but as a slightly steeper power of $k_1$. As pointed out by
\cite{delalamo&jimenez09}, observed deviations from $k_{1}^{-1}$ could
result from a failure of the Taylor hypothesis, a point which we do
not dispute. However, we show in this section that slightly steeper
powers of $k_1$ can also arise because of intermittent fluctuations of
the wall shear stress, as observed for example by
\cite{alfredssonetal88} and \cite{orlu&schlatter11}.

%The argument by Perry, Henbest \& Chong (1986) leading to
%$E_{11}(k_{1},y) \sim u_{\tau}^{2} k_{1}^{-1}$ in $1/\delta \ll k_{1}
%\ll 1/y$ is based on the assumptions that for wavenumbers $k_1 \ll
%1/y$, $E_{11}(k_{1},y) = u_{\tau}^{2} \delta g_{o}( k_{1}\delta )$,
%whereas for wavenumbers $1/\delta \ll k_{1} \ll 1/\eta$,
%$E_{11}(k_{1},y) = u_{\tau}^{2} y g_{i}( k_{1}y)$ (in terms of outer
%and inner dimensionless functions $g_o$ and $g_i$). In the overlap
%range of wavenumbers $1/\delta \ll k_{1} \ll 1/y$, $E_{11}(k_{1},y) =
%u_{\tau}^{2} \delta g_{o}( k_{1}\delta ) = u_{\tau}^{2} y g_{i}(
%k_{1}y)$. Hence $\delta g_{o}( k_{1}\delta ) = y g_{i}( k_{1}y) =
%k_{1}^{-1}$. This overlap range exists if $Re_{\tau} \equiv
%\delta/\delta_{\nu} \gg 1$ and $\delta_{\nu} \ll y \ll \delta$.

One way to argue, in the region $\delta_{\nu}\ll y\ll \delta$, that
$E_{11}(k_{1},y) \sim u_{\tau}^{2} k_{1}^{-1}$ in the wavenumber range
$1/\delta \ll y \ll 1/y$ is by hypothesizing that the attached eddies
dominate the spectrum in that range independently of $y$ and that
these eddies are themselves dominated by the wall shear stress,
i.e. the skin friction, at the wall. Hence $E_{11}(k_{1},y)$ can only
depend on $u_{\tau}^{2}$ and $k_{1}$ in the region $\delta_{\nu}\ll
y\ll \delta$, which implies that $E_{11}(k_{1},y) \sim u_{\tau}^{2}
k_{1}^{-1}$.

We now show how this argument can be modified to take into account the
intermittency in the wall shear stress. To do this we adopt the way
that \cite{kolmogorov62} took into account the inertial-range
intermittency of kinetic energy dissipation in homogeneous turbulence
and adapt it to the intermittency of wall shear stress in wall
turbulence. We therefore define the scale-dependent filter averages
\begin{equation}
u_{*}^{2}(x,r,t) = {1 \over
2r} \int_{x-r}^{x+r} \nu \left. {du\over dy} \right \vert_{wall}(x,t) \, dx. 
\label{eq:5.1}
\end{equation}

Following Kolmogorov's (1962) approach we assume that the statistics
of $u_{*}^{2}(x,r,t)$ are lognormal at scales $r$ large enough for
$u_{*}^{2}(x,r,t)$ to be reasonably presumed positive. It may be
reasonable to assume scales $r$ much larger than $y$ to be such scales
if $\delta_{\nu} \ll y \ll \delta$. For such scales we therefore
define $\xi_{r} \equiv \ln (u_{*}^{2}/u_{\tau}^{2})$ and assume
$\xi_r$ to be a gaussian-distributed random variable, i.e. its PDF is
\begin{equation}
P(\xi_{r}) = {1\over \sqrt{2\pi} \sigma_{r}} e^{-(\xi_{r} - m_{r})^{2}/2\sigma_{r}^{2}}
\label{eq:5.2}
\end{equation}
The constraint $<u_{*}^{2}(x,r,t)> = u_{\tau}^{2}$ implies $m_{r} =
-\sigma_{r}^{2}/2$. The exact form of this PDF does not really matter
as we are only concerned with low order moments.

We now hypothesize that, in the appropriate Townsend-Perry attached
eddy range of wavenumbers, the average of $(u'(x+r,y)-u'(x,y))^{2}$
conditioned on $u_{*}^{2}(x,r,t)$ taking a certain value depends only
on that value and $r$ ($u'$ is the streamwise fluctuating turbulence
velocity component). By dimensional analysis the dependence on $r$
drops out, and as the structure function $<(u'(x+r,y)-u'(x,y))^{2}>$
is the average over all these conditional averages, we are left with
$<(u'(x+r,y)-u'(x,y))^{2}> \sim <u_{*}^{2}(x,r,t)>$. Using (\ref{eq:5.2}) to
calculate this average, we obtain
\begin{equation}
<(u'(x+r,y)-u'(x,y))^{2}> \sim u_{\tau}^{2} \int d\xi {e^{\xi}\over
  \sqrt{2\pi} \sigma_{r}} e^{-(\xi - m_{r})^{2}/2\sigma_{r}^{2}} \sim
u_{\tau}^{2} e^{-\sigma_{r}^{2}/9}.
\label{eq:5.3}
\end{equation}
A logarithmic dependence of $\sigma_{r}^{2}$ or $r$, for example
$\sigma_{r}^{2} = const + 9\mu \ln (\delta /r)$ where $\mu >0$,
returns $<(u'(x+r,y)-u'(x,y))^{2}> \sim u_{\tau}^{2} (r/\delta)^{\mu}$, i.e. 
\begin{equation}
E_{11} (k_{1},y)\sim u_{\tau}^{2} \delta (k_{1}\delta)^{-1-\mu}.
\label{eq:5.4}
\end{equation}

This demonstrates that the attached eddy hypothesis suitably modified
to take into account the intermittent fluctuations of the wall shear
stress can lead to spectra that are slightly steeper than
$k_{1}^{-1}$. The statistics of the intermittently fluctuating wall
shear stress can therefore have some bearing on energy spectra and, in
turn, on vertical profiles of the turbulent kinetic energy. One can
readily see that replacement of $E_{11} (k_{1},y) \approx C_{0}
u_{\tau}^{2} k_{1}^{-1}$ by $E_{11} (k_{1},y) \approx C_{0}
u_{\tau}^{2} \delta (k_{1}\delta)^{-1-\mu}$ in range (ii) of the
\cite{perryetal86} model (section 3) and in range (iii) of our
modified model in section 4 would lead to profiles such as (\ref{eq:4.5}) and
(\ref{eq:4.9}) where the $\ln(\delta/y)$ terms would be replaced by weak power
laws of $y/\delta$. However, for very small exponents $\mu$ this
difference would be very hard to detect experimentally.

\section{The mean flow profile}

As already noted by \cite{townsend76}, the attached eddy hypothesis is
incompatible with the assumption that ${d \overline{u} \over dy}$ is
independent of $\delta$. This assumption is required to argue that ${d
  \overline{u} \over dy}$ depends only on $y$ and $u_{\tau}$ in the
range $\delta_{\nu} \ll y \ll \delta$. As $Re_{\tau}\to \infty$ an
intermediate layer $\delta_{\nu}\ll y \ll \delta$ does emerge,
however, where something may nevertheless be independent of $\nu$ and
$\delta$. \cite{dallasetal09} presented evidence from DNS of turbulent
channel flow which shows that the eddy turnover time $\tau \equiv
E/\epsilon$ (where $E$ is the total turbulent kinetic energy) is
proportional to $y/u_{\tau}$ in the range $\delta_{\nu} \ll y \ll
\delta$ for a variety of, admitedly very moderate, values of
$Re_{\tau}$.

Here we make the reasonable extrapolation that the observation of
\cite{dallasetal09} is not limited to moderate Reynolds numbers and
that $\tau$ is independent of $\nu$ and $\delta$ at all large enough
Reynolds numbers. Hence, $\tau \sim {y\over u_{\tau}}$ in the range
$\delta_{\nu} \ll y \ll \delta$ for turbulent pipe/channel flows.

Following \cite{townsend76} we also assume local balance between
production and dissipation, i.e. $-<u'v'> {d\overline{u}\over dy}
\approx \epsilon = E/\tau$, but only in a region $y_{P\epsilon} < y
\ll \delta$ where $\delta_{\nu} \ll y_{P\epsilon}$. Making use of the
well-known axial momentum balance in turbulent pipe/channel flow, see
\cite{pope00}, 
\begin{equation}
\nu {d\over dy} \overline{u} - <u'v'> = u_{\tau}^{2}(1-y/\delta),
\label{eq:6.1}
\end{equation}
and introducing the constant $C_s$ in $\tau \approx C_{s} {y\over
  u_{\tau}}$, we are led to
%\begin{equation}
%(u_{\tau}^{2}(1-y/\delta) - \nu
%    {d\overline{u}\over dy}){d\overline{u}\over dy} \approx
%    C_{s} {u_{\tau}\over y} E
%\end{equation}
%in the region $y_{P\epsilon} < y \ll \delta$. This equation can also
%be written as
\begin{equation}
(1-y/\delta - {d\overline{u}_{+}\over dy_{+}}){d\overline{u}_{+}\over
    d \ln y_{+}} \approx C_{s} E/u_{\tau}^{2} =C_{s} E_{+}.
\label{eq:6.2}
\end{equation}
in the region $y_{P\epsilon} < y \ll \delta$. 

If a $y$-region exists where $E_{+}$ is constant with respect to $y$
and $Re_{\tau}$ and if the Reynolds number is high enough for
$(1-y/\delta - {d\overline{u}_{+}\over dy_{+}})$ to be approximately
1, then (\ref{eq:6.2}) is just the well-known log law. However, we know from
the Townsend-Perry attached eddy
model and also from this paper's modified such model that $E_{+}
\approx M_{0} + M_{1} \ln(\delta/y)$ in the range $y_{*} < y \ll
\delta$ where $M_{0}$ and $M_1$ are constants different from 
$C_{\infty}$ and $C_0$ in (\ref{eq:3.2}) because one needs to also take into
account ${1\over 2} \overline{w'^{2}}(y)/u_{\tau}^{2}$ and ${1\over 2}
\overline{v'^{2}}(y)/u_{\tau}^{2}$. Hence the first prediction of our
approach based on $\tau \approx C_{s} {y\over u_{\tau}}$ and $-<u'v'>
{d\overline{u}\over dy} \approx \epsilon$ is that the left hand side
of (\ref{eq:6.2}) is approximately equal to $C_{s} M_{0} + C_{s} M_{1}
\ln(\delta/y)$ in $y_{*} < y \ll \delta$.

If $E_+$ has an outer peak at the same $y=y_{peak}$ location as
${1\over 2} \overline{u'^{2}}(y)/u_{\tau}^{2}$ and if $y_{P\epsilon} <
y_{peak}$ then the second prediction of our approach is that the left
hand side of (\ref{eq:6.2}) has an outer peak at $y=y_{peak}$.

\begin{figure}
\centering 
\vspace{1cm}
\includegraphics[width=0.8\columnwidth]{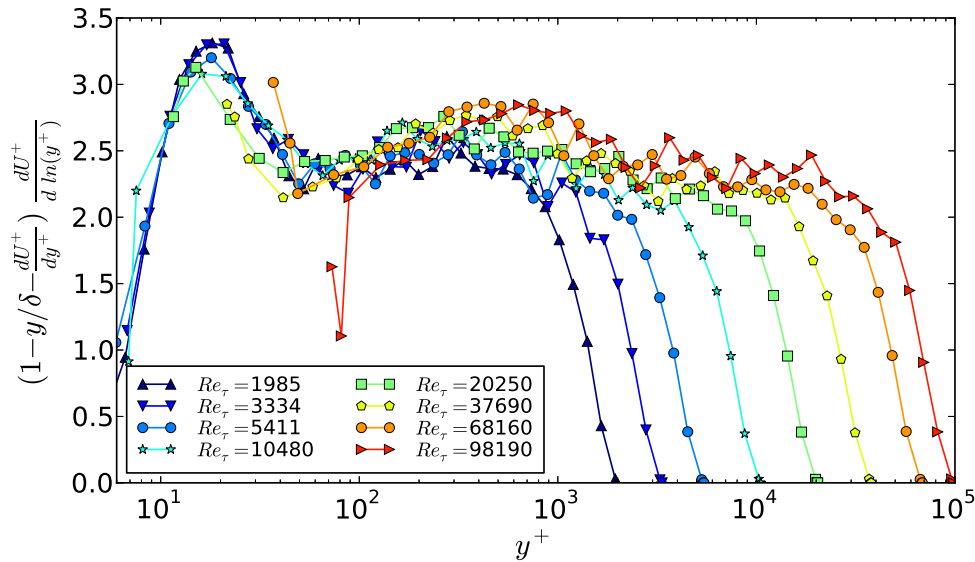}
\caption{Linear-logarithmic plot of $(1-y/\delta -
  {d\overline{u}_{+}\over dy_{+}}){d\overline{u}_{+}\over d \ln
    y_{+}}$ versus $y_+$ for different values of $Re_{\tau}$ obtained
  from the NSTAP Superpipe mean flow data of \cite{hultmarketal12,
    hultmarketal13}.}
\label{fig:6}
\end{figure}

%One can expect a peak in $E_+$ at the same location as the peak in
%${1\over 2} \overline{u'^{2}}(y)/u_{\tau}^{2}$ because, compared to
%the wall-normal and transverse contributions, ${1\over 2}
%\overline{u'^{2}}(y)$ dominates $E_+$.

\begin{figure}
\centering 
\vspace{1cm}
\includegraphics[width=0.8\columnwidth]{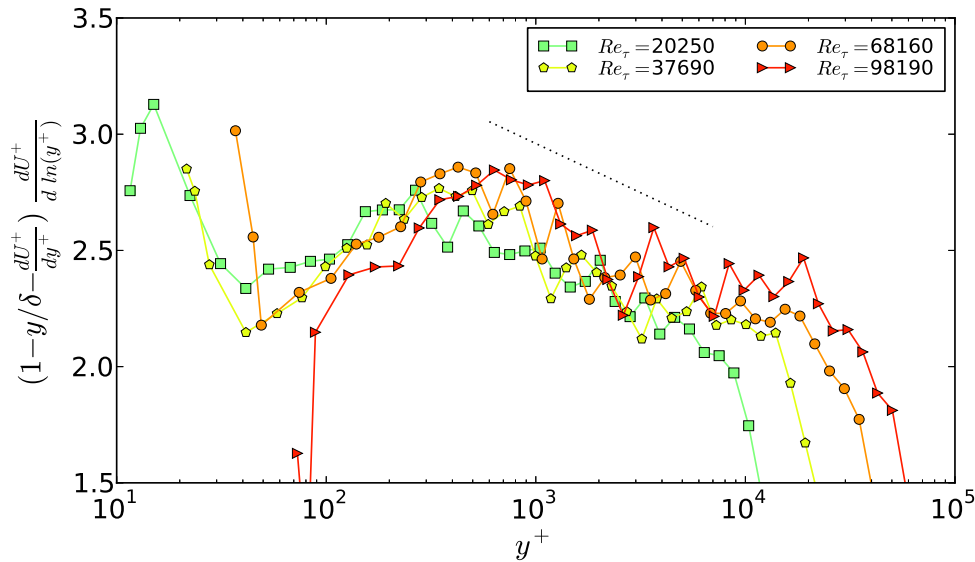}
\caption{Blow up of figure 6 for the four highest Reynolds numbers
  with a superposed dotted line suggesting logarithmic dependence of
  $(1-y/\delta - {d\overline{u}_{+}\over
    dy_{+}}){d\overline{u}_{+}\over d \ln y_{+}}$ on $y$ at the right
  of the peak.}
\label{fig:7}
\end{figure}

Figure \ref{fig:6} is a plot of the left hand side of (\ref{eq:6.2}) based on the NSTAP
Superpipe data of \cite{hultmarketal12, hultmarketal13}. This plot
suggests that there is indeed an outer peak in the functional
dependence on $y$ of the left hand side of (\ref{eq:6.2}). It is also not
inconsistent with the prediction that the left hand side of (\ref{eq:6.2}) is a
logarithmically decreasing function of $y$ for much of the region
where $y$ is greater than the location of this outer peak. Figure \ref{fig:7}
shows this left hand side for the higher $Re_{\tau}$ NSTAP Superpipe
data ($Re_{\tau}$ between $20\,000$ and $100\,000$)
There is no evidence that the left-hand side of (\ref{eq:6.2}) decreases
logarithmically with $y$ for the lower Reynolds numbers in figure \ref{fig:6},
in agreement with (\ref{eq:6.2}) and figures \ref{fig:1} and \ref{fig:3} which show that there is
no such logarithmic decrease in ${1\over 2}
\overline{u'^{2}}(y)/u_{\tau}^{2}$ either at $Re_{\tau} <
10\,000$. However such a $y$ dependence is not inconsistent with much of
the $y$-dependence for the $Re_{\tau} > 20\,000$ data at the right of
the outer peak in figure \ref{fig:7}.

\begin{figure}
\centering 
\vspace{1cm}
\begin{minipage}{0.49\columnwidth}
\includegraphics[width=\columnwidth]{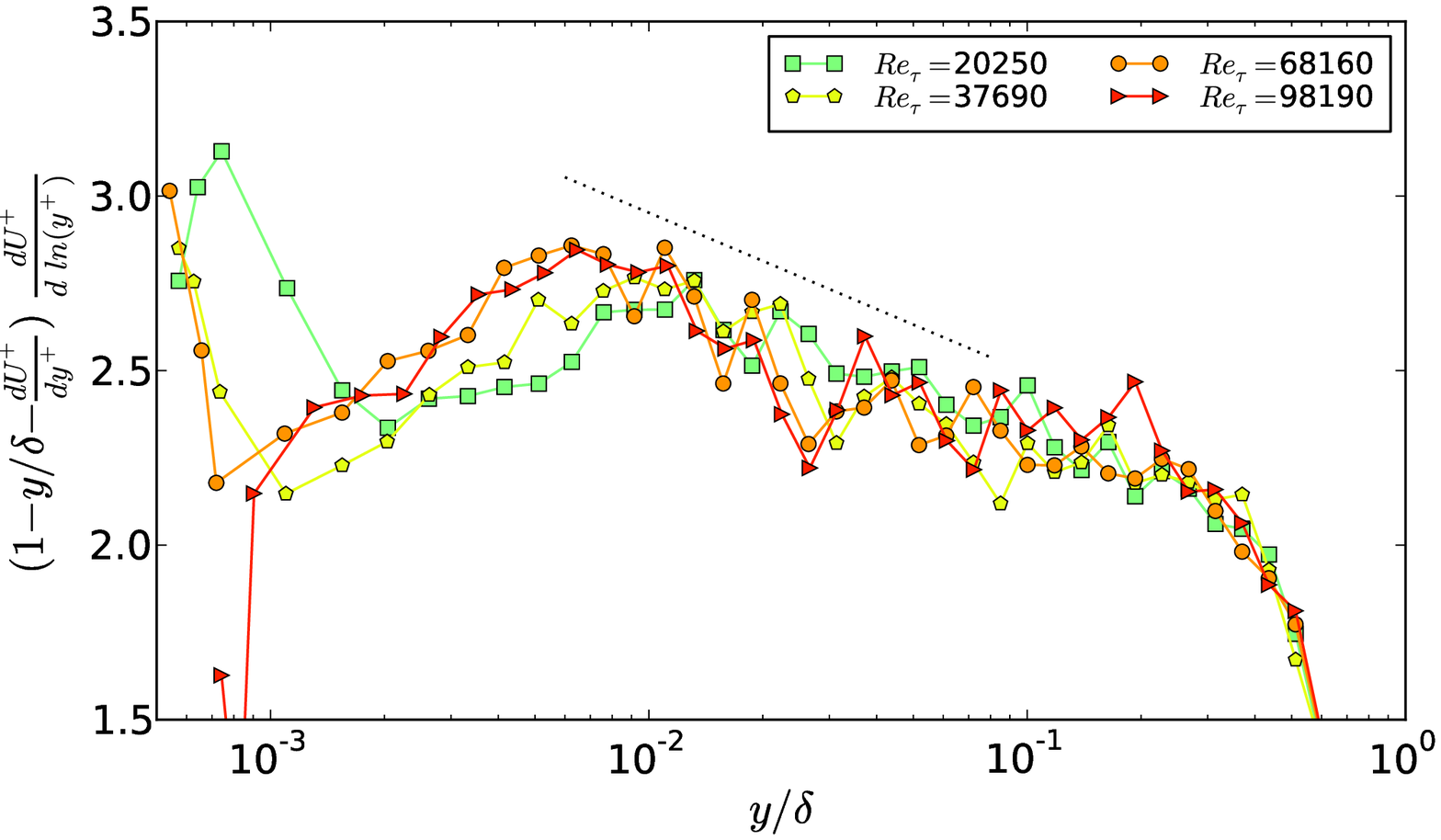}
\end{minipage}
\hfill
\begin{minipage}{0.49\columnwidth}
\includegraphics[width=\columnwidth]{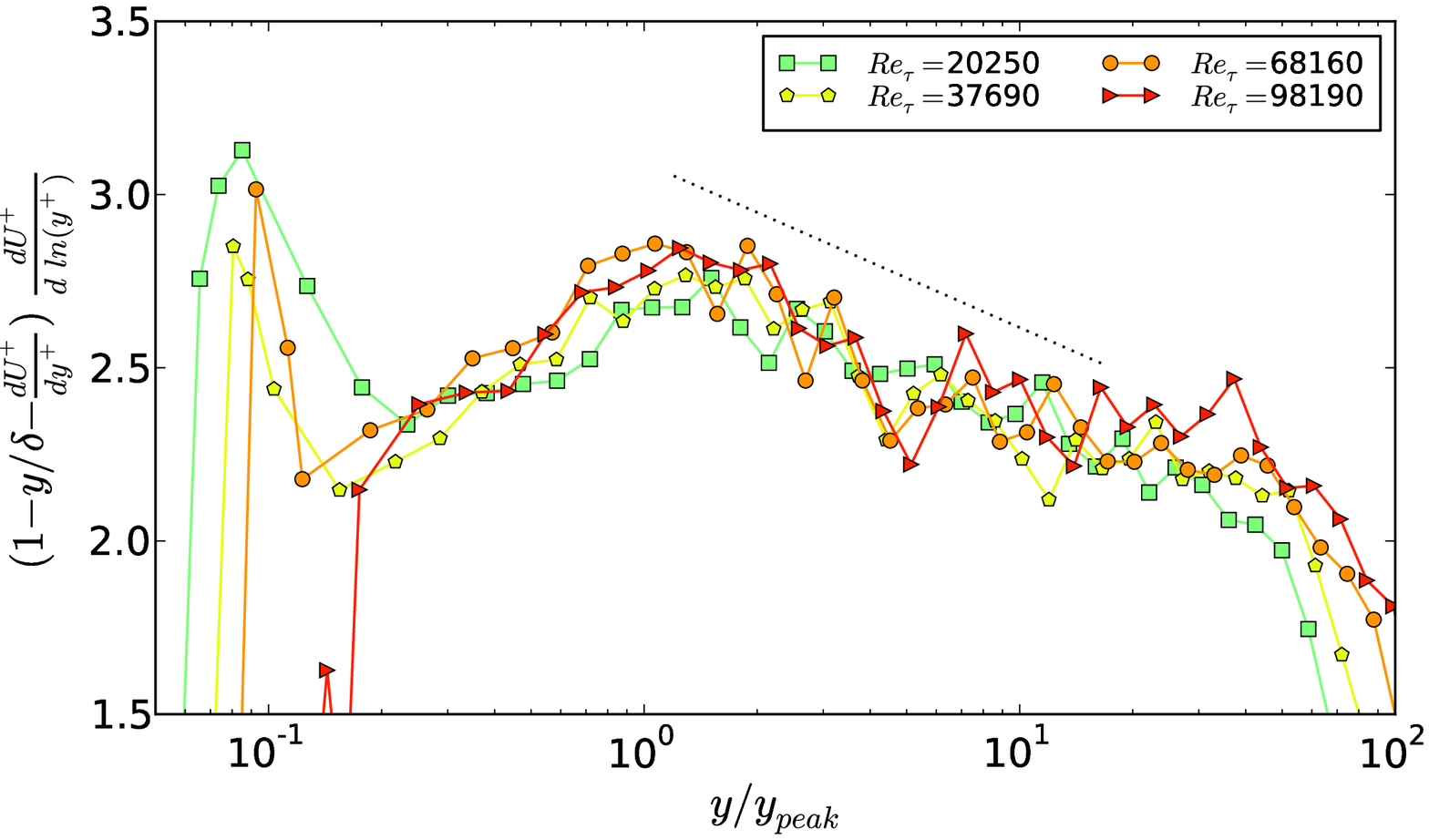}
\end{minipage}
\caption{Blow ups of figure \ref{fig:6} for the four highest Reynolds numbers
  plotted versus $y/\delta$ (left) and versus $y/y_{peak}$ (right)
  where $y_{peak} = 0.23 \delta_{\nu} Re_{\tau}^{0.67}$ is the fit by
  \cite{hultmarketal12} of the location of the outer peak in the
  streamwise turbulent energy plotted in figures \ref{fig:1} and \ref{fig:3}. The
  superposed dotted line suggests a logarithmic dependence of
  $(1-y/\delta - {d\overline{u}_{+}\over
    dy_{+}}){d\overline{u}_{+}\over d \ln y_{+}}$ on $y/\delta$ at the
  right of the peak.}
\label{fig:8}
\end{figure}

In figure \ref{fig:8} we replot the high $Re_{\tau}$ data of figure \ref{fig:7} but as
functions of $y/\delta$ in one plot and of $y/y_{peak}$ in the
other. These plots demonstrate that the position of the outer peak in
the left-hand side of (\ref{eq:6.2}) is the same as the position of the outer
peak in ${1\over 2} \overline{u'^{2}}(y)/u_{\tau}^{2}$. And they also
demonstrate that the left hand side of (\ref{eq:6.2}), if indeed
logarithmically decreasing, is approximately equal to $C_{s} M_{0} +
C_{s} M_{1} \ln(\delta/y)$ in $y_{*} < y \ll \delta$ (though the data
in our disposal do not permit us to check that the constants $C_{s}
M_{0}$ and $C_{s} M_{1}$ are indeed the products of $C_s$ with $M_0$
and $M_1$ respectively).

\begin{figure}
\centering 
\begin{minipage}{0.49\columnwidth}
\includegraphics[width=\columnwidth]{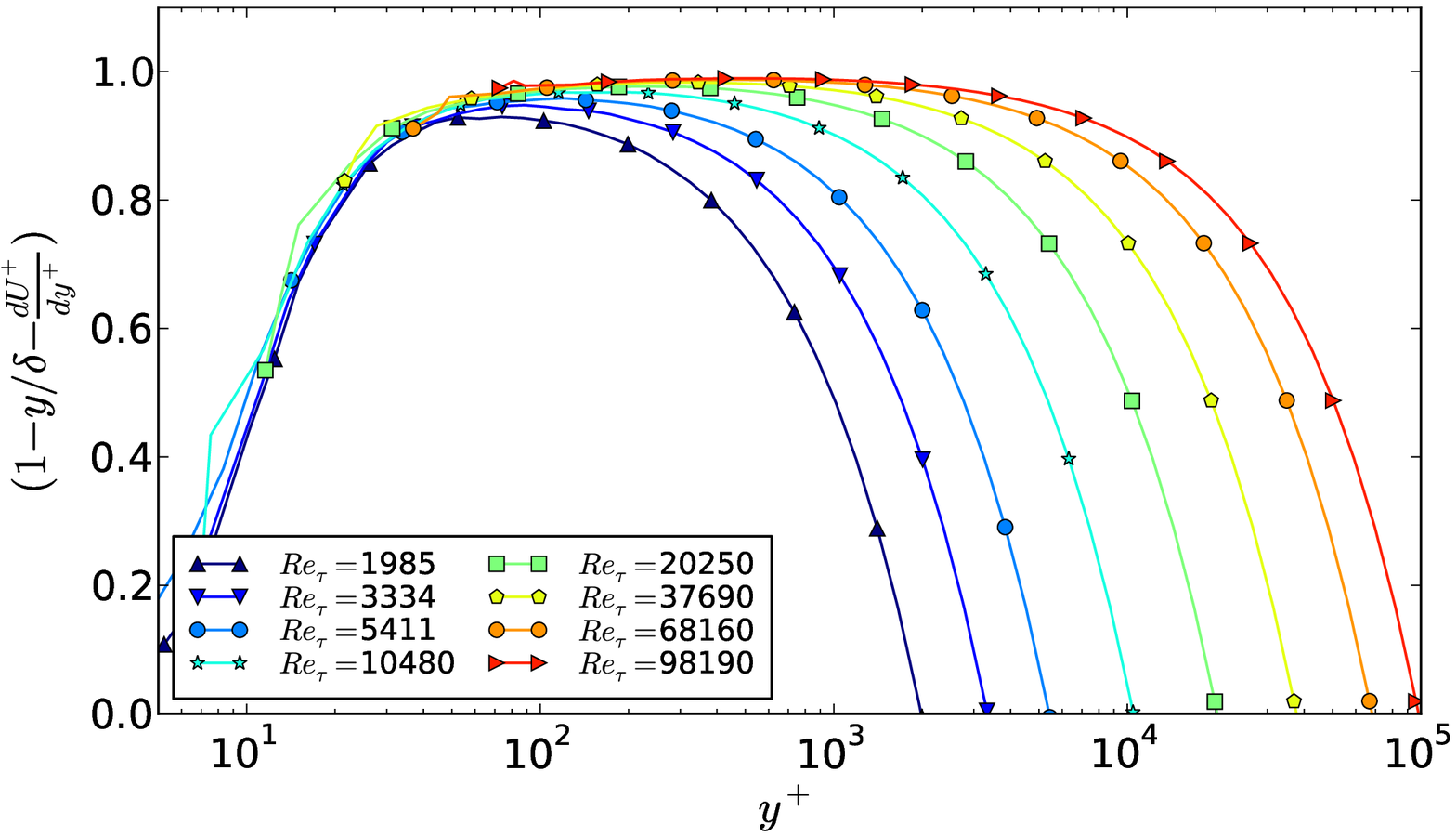}
\end{minipage}
\hfill
\begin{minipage}{0.49\columnwidth}
\includegraphics[width=\columnwidth]{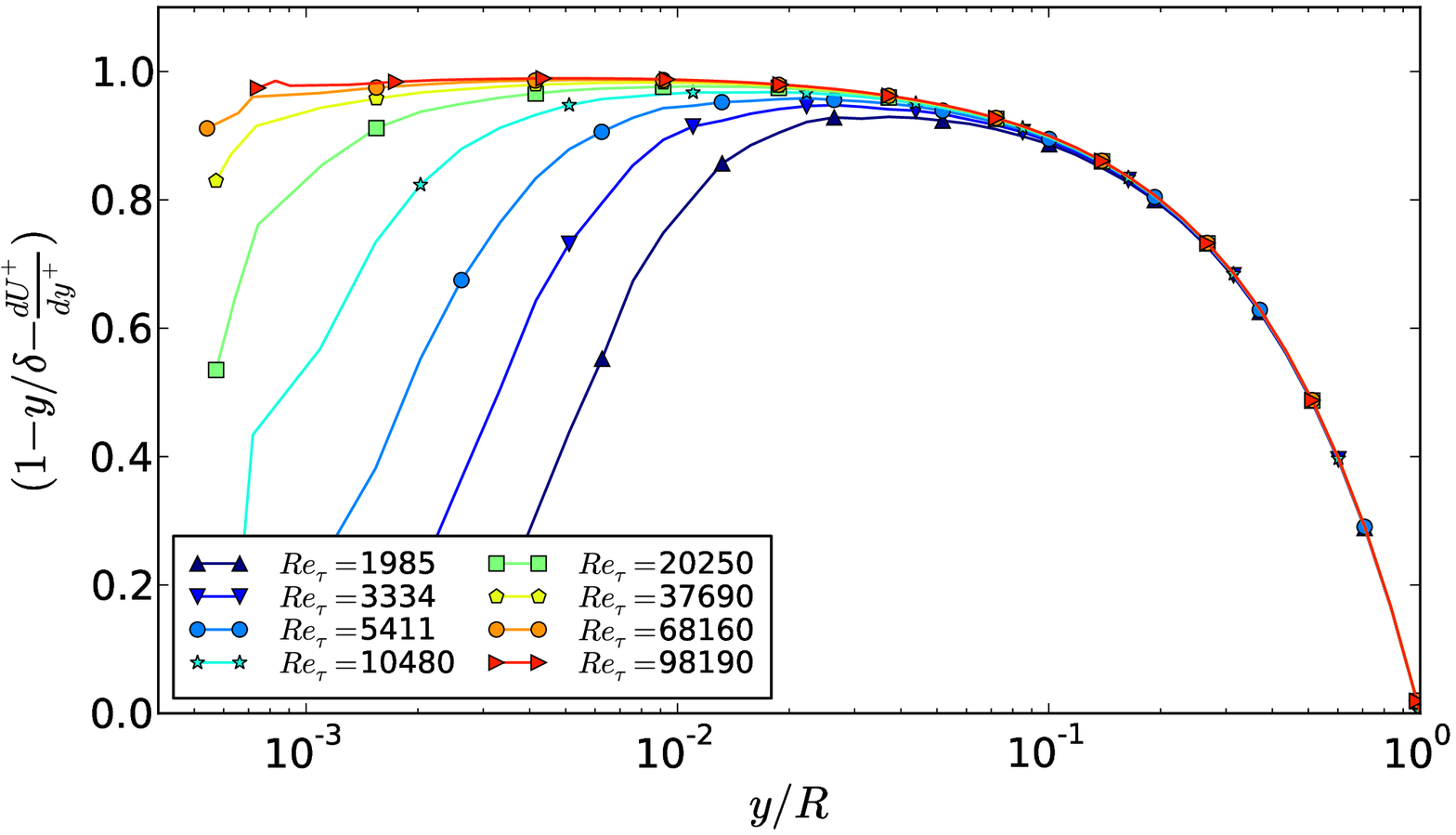}
\end{minipage}
\caption{Normalised Reynolds stress $-<u'v'>/u_{\tau}^{2}$ calculated
  from the NSTAP Superpipe mean flow data of \cite{hultmarketal12,
    hultmarketal13} as $(1-y/\delta - {d\overline{u}_{+}\over
    dy_{+}})$ (for turbulent pipe flow) versus $y_+$ (lefft) and
  versus $y/\delta$ (right). $Re_{\tau}$ ranges from about $2\,000$ to
  about $100\,000$.}
\label{fig:9}
\end{figure}

In figure \ref{fig:9} we use the NSTAP Superpipe data to plot $ (1-y/\delta -
{d\overline{u}_{+}\over dy_{+}})$ as a function of $y/\delta$ in one
case and $y_+$ in the other. As these are pipe data, the plots in
figure \ref{fig:9} are effectively plots of the normalised Reynolds stress
$-<u'v'>/u_{\tau}^{2}$.  It is clear that $-<u'v'> \approx
u_{\tau}^{2}$ only if $Re_{\tau} > 40\,000$ and for distances from the
wall such that $100< y_{+}$ and $y/\delta < 0.01$. At values of $y$
larger than $\delta/10$ the normalised Reynolds stress decreases
abruptly towards 0 which explains why the left hand side of (\ref{eq:6.2}) does
the same in figures \ref{fig:6} to \ref{fig:8} at these values of $y$.

Figure \ref{fig:9} makes it clear that equation (\ref{eq:6.2}) simplifies to
\begin{equation}
{d\overline{u}_{+}\over d \ln y_{+}} \approx C_{s} E_{+}
\label{eq:6.3}
\end{equation}
in turbulent pipe flow only if $Re_{\tau}>40\,000$ and only in the range
$100\delta_{\nu} < y < \delta/100$. Using the attached eddy model's
$E_{+} \approx M_{0} + M_{1} \ln(\delta/y)$ in the range $y_{*} < y
\ll \delta$ we obtain the following asymptotic form of the mean flow
profile in $y_{*}< y < 0.01\delta$ (as $y_*$ is larger than $100\delta_{\nu}$):
\begin{equation}
\overline{u}_{+} \approx C_{s}M_{0} \ln(y/\delta) -
  {C_{s}M_{1}\over 2} [\ln(y/\delta)]^{2}+ M_{2}
\label{eq:6.4}
\end{equation}
in terms of an extra integration constant $M_2$. We stress again the
limited $y$-range of validity of this high Reynolds number mean flow
profile (to the right of the outer peak) and that it can only be
expected at $Re_{\tau} > 40\,000$. 

As shown in section 5, $E_{+} \approx M_{0} + M_{1} \ln(\delta/y)$ and
therefore also (\ref{eq:6.4}) are based on the additional assumption that any
intermittency which might exist in the fluctuating wall shear stress
is of such a nature that the Townsend-Perry spectral scalings $E_{11}
(k_{1},y)\sim u_{\tau}^{2} k_{1}^{-1}$ remain intact. Otherwise one
can expect power laws of $y/\delta$ instead of logarithms of
$y/\delta$ in the formula for the mean flow profile (\ref{eq:6.4}).

We close this section with a comment on the mesolayer, a concept
introduced by \cite{long&chen81} and most recently discussed by
\cite{vallikivietal14} who also provide a list of relevant
references. In the present paper, profiles have been obtained for
$\overline{u'^{2}}(y)$ in the range $\delta_{\nu} \ll y \ll \delta$
and for $\overline{u} (y)$ in the range $y_{P\epsilon} <y < 0.01
\delta$ where production has been assumed to balance dissipation.
%Production is known to greatly exceed dissipation at $y_{+} \approx
%15$ where there is an inner peak both in the turbulence energy
%(figures 1,3) and in the mean flow gradient profiles (figure 7).
\cite{george&castillo97} argued that the mesolayer is a region from
$y_{+} \simeq 30$ to $y_{+} \simeq 300$ where, owing to low turbulent Reynolds
number $y_+$ values, the dissipation does not have its high Reynolds
number scaling and the Kolmogorov range (iv) of our spectral model in
section 4 is effectively absent. This has no bearing on our turbulent
kinetic energy calculations of sections 4 and 5 because the energy in
the Kolmogorov range (iv) is small compared to the other ranges and
the outer peak comes from the new small wavenumber range (ii). (In
fact it is easy to check that the Kolmogorov range in the
Townsend-Perry model cannot, by itself, lead to an outer turbulent
energy peak.) However, it might be that we cannot use the scaling
$\tau \sim y/u_{\tau}$ at $y_{+} \lesssim 300$ and that our approach for
obtaining the mean flow gradient profile might therefore be valid only
in the region $\max (300\delta_{\nu}, y_{P\epsilon}) < y \ll
0.01\delta$. Note that the value of $y_{peak}$ in the Princeton NSTAP
data is about $300\delta_{\nu}$ at $Re_{\tau} \approx 40\,000$ and about
$500\delta_{\nu}$ at $Re_{\tau} \approx 100\,000$, which means that the
mesolayer is indeed under $y_{peak}$ for $Re_{\tau} >40\,000$. The
prediction that the mean flow gradient has an outer peak at the same
distance from the wall where the turbulent kinetic energy has an outer
peak has been based on the assumption that $y_{P\epsilon} <
y_{peak}$. The region where production and dissipation balance and
where turbulent transport has negligible effects may or may not be
expected to have an overlap with the mesolayer. The task of working
out the scalings of $y_{P\epsilon}$ and how it compares with
$300\delta_{\nu}$ must be left for a future study which will have the
means to address these questions.

\section{Conclusion}

In way of conclusion we list the main points made in this paper.

1. For the Townsend-Perry $k_{1}^{-1}$ spectrum to be viable, i.e. to
be compatible with a realistic integral scale dependence on $y$, we
need to add to the \cite{perryetal86} spectral model an extra
wavenumber range at wavenumbers smaller than those where $E_{11}
(k_{1},y) \sim u_{\tau}^{2} k_{1}^{-1}$.

2. Simple modelling of this range (see section 4) implies the
existence of an outer peak in the streamwise turbulence kinetic energy
at a $y$-position $y_{peak}$ which grows with respect to
$\delta_{\nu}$ and decreases with respect to $\delta$ as $Re_{\tau}$
increases. The streamwise kinetic energy at that peak grows
logarithmically with $Re_{\tau}$.

3. The functional form which results from our modified Townsend-Perry
model and which may be useful as a starting point in future
investigations is the following: in the range $\delta_{\nu} \ll y <
y_{*} \sim \delta Re_{\tau}^{-1/3}$
\begin{equation}
{1\over 2} \overline{u'^{2}}(y)/u_{\tau}^{2} \approx C_{s0} -
C_{s1}\ln (\delta/y) -C_{s2} (y/\delta)^{d} Re_{\tau}^{d/3}
\label{eq:7.1}
\end{equation} 
where all the constants are independent of $y$, $\delta$, $\nu$ and
$Re_{\tau}$ except for $C_{s0}$ which may be a logarithmically
increasing function of $Re_{\tau}$; in the range $y_{*} < y
\ll \delta$
\begin{equation}
{1\over 2} \overline{u'^{2}}(y)/u_{\tau}^{2} \approx C_{3} + C_{4}\ln
  (\delta/y)
\label{eq:7.2}
\end{equation}
as predicted by \cite{townsend76} and \cite{perryetal86}. 

4. The very high $Re_{\tau}$ Princeton Superpipe NSTAP data used here
and the turbulent channel flow DNS of \cite{dallasetal09}
support the view that it is the eddy turnover time $\tau \equiv
E/\epsilon$ that is independent of $\nu$ and $\delta$ in the range
$\delta_{\nu} \ll y \ll \delta$ rather than the mean flow
gradient. This implies $\tau \sim y/u_{\tau}$ in that range, a
relation which can serve as a unifying principle across Reynolds
numbers in turbulent pipe/channel flows. Of course, further research
is needed to fully establish such a unifying principle.

5. The mean flow profile and scalings can be obtained from $\tau \sim
y/u_{\tau}$ if enough is known about the production-dissipation
balance/imbalance. Here we have assumed that production and
dissipation balance in a range $y_{P\epsilon} < y \ll \delta$ where
$y_{P\epsilon}$ 
%scales as $y_{peak}$ and may be smaller than
is smaller than $y_{peak}$. Due to this balance, a profile for $E_{+}$
similar to that of $\overline{u'^{2}}/u_{\tau}^{2}$ and $-<u'v'>
\approx u_{\tau}^{2}$ imply that ${d\overline{u}_{+}\over d \ln
  y_{+}}$ (i) has an outer peak at the same position $y=y_{peak}$
where $\overline{u'^{2}}/u_{\tau}^{2}$ has an outer peak, and (ii)
decreases with distance from the wall as a function of $\ln
(\delta/y)$ where $y_{*} < y \ll \delta$. The very high $Re_{\tau}$
NSTAP Princeton Superpipe data show clear evidence of both these
features.

6. The NSTAP Princeton Superpipe data also show that the
Reynolds stress $<u'v'>$ is approximately equal to $-u_{\tau}^{2}$
only if $Re_{\tau} > 40\,000$ and for distances from the wall such that
$100< y_{+}$, $y/\delta < 0.01$. The balance $-<u'v'> {d\overline{u}
  \over dy}\approx \epsilon$ and the kinetic energy profile $E_{+}
\approx M_{0}+ M_{1}\ln (\delta/y)$ (where $M_0$ and $M_1$ are
dimensionless constants) in $y_{*} \ll y \ll \delta$ therefore imply
in terms of an integration constant $M_2$ that
\begin{equation}
\overline{u}_{+} \approx C_{s}M_{0} \ln(y/\delta) -
  {C_{s}M_{1}\over 2} [\ln(y/\delta)]^{2}+ M_{2}
\end{equation}
in $y_{*}< y < 0.01\delta$ provided that $Re_{\tau} > 40\,000$. This is
the modified log-law of the wall.

\vskip 1truecm

\subsection*{Acknowledgements}
We are very grateful to Dr M. Vallikivi and Professor A. J. Smits for
kindly providing us with their NSTAP Superpipe data (first published
in \cite{hultmarketal12, hultmarketal13}) and with energy spectra from
the same measurements. This work was supported by Campus International
pour la S\'ecurit\'e et l'Intermodalit\'e des Transports, la R\'egion
Nord-Pas-de-Calais, l'Union Europ\'eenne, la Direction de la
Recherche, Enseignement Sup\'erieur, Sant\'e et Technologies de
l'Information et de la Communication et le Centre National de la
Recherche Scientifique.  JCV acknowledges the support of an ERC
Advanced Grant (2013-2018).

\end{document}